\documentclass[aps,prl,twocolumn,showpacs,superscriptaddress,groupedaddress]{revtex4-1} 
\usepackage{graphicx}
\usepackage{subfigure}
\usepackage{bm}
\usepackage{latexsym,amsthm,amssymb}
\usepackage{booktabs,threeparttable}
\usepackage{makeidx}
\usepackage{amsmath}
\usepackage{amsfonts}
\usepackage{amssymb}
\usepackage{setspace}
\usepackage{slashed}

\begin{document}
%%%%%%%%%%%%%%%%%%%%%%%%%%%%%%%%%%%%%%%%%%%%%%%%%%%%%
\title{Natural Negative-Refractive-Index Materials}
\author{Chih-Yu Chen$^{1, 3}$, Ming-Chien Hsu$^{2}$, C.D. Hu$^{1}$, Yeu Chung Lin$^{1}$}
\affiliation{1. National Taiwan University, Taipei, Taiwan\\ 2. National Sun Yat-Sen University, Kaohsiung, Taiwan \\3. Chung Yuan Christian University, Taoyuan, Taiwan} 
\date{\today }

\begin{abstract}
Our calculation shows that negative refractive index (NRI), which was known to exist only in metamaterials in the past, can be found in Dirac semimetals (DSM). Electrons in DSM have zero effective mass and hence the system carries no nominal energy scale. Therefore, unlike those of ordinary materials, the electromagnetic responses of the electrons in DSM will not be overwhelmed by the physical effects related to electron mass. NRI is induced by the combination of the quantum effect of vacuum polarization and its finite temperature correction which is proportional to $T^4$ at low temperature. It is a phenomenon of resonance between the incident light and the unique structure of Dirac cones, which allows numerous states to participate in electron-hole pair production excited by the incident light with similar dispersion relation to that of Dirac cones. NRI phenomenon of DSM manifests in an extensive range of photon frequency and wave number and can be observed around giga Hertz range at temperature slightly higher than room temperature.
\end{abstract}
\maketitle
\begin{figure}[t]
\centering
\subfigure[ ]{
\includegraphics[scale=0.43]{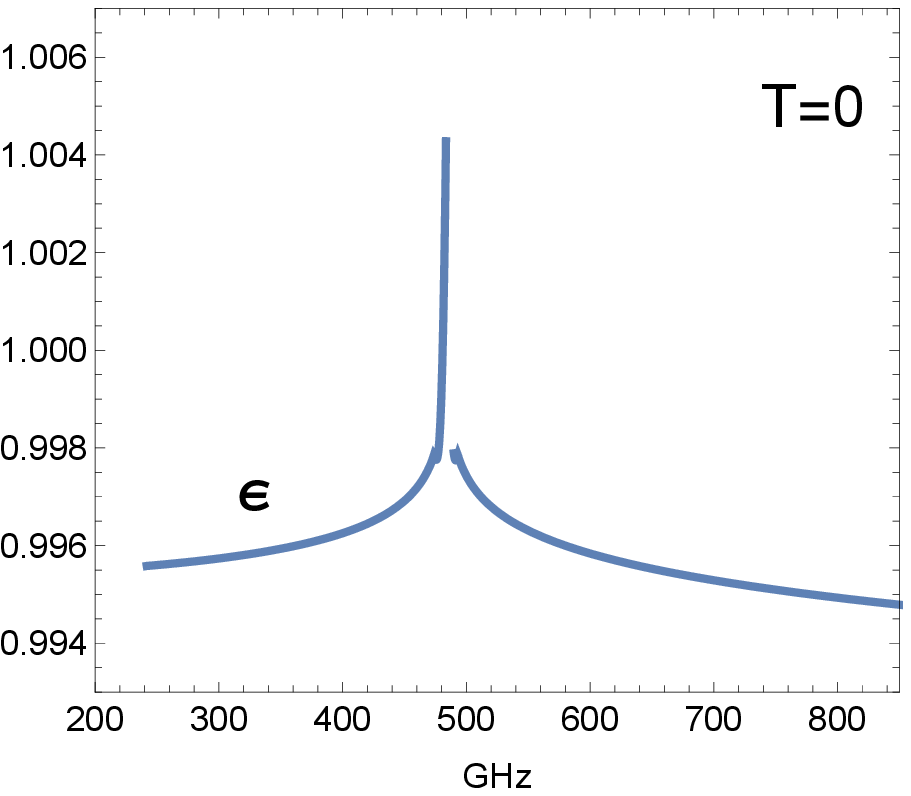}}
\hspace{0.1cm}
\subfigure[ ]{
\includegraphics[scale=0.48]{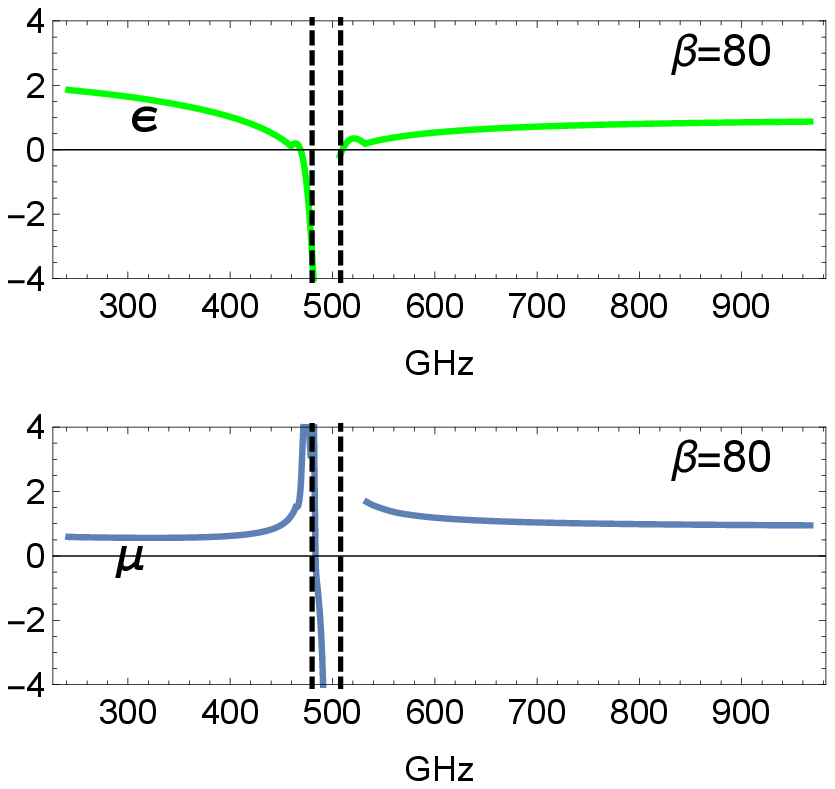}}
\hspace{0.1cm}
\subfigure[ ]{
\includegraphics[scale=0.48]{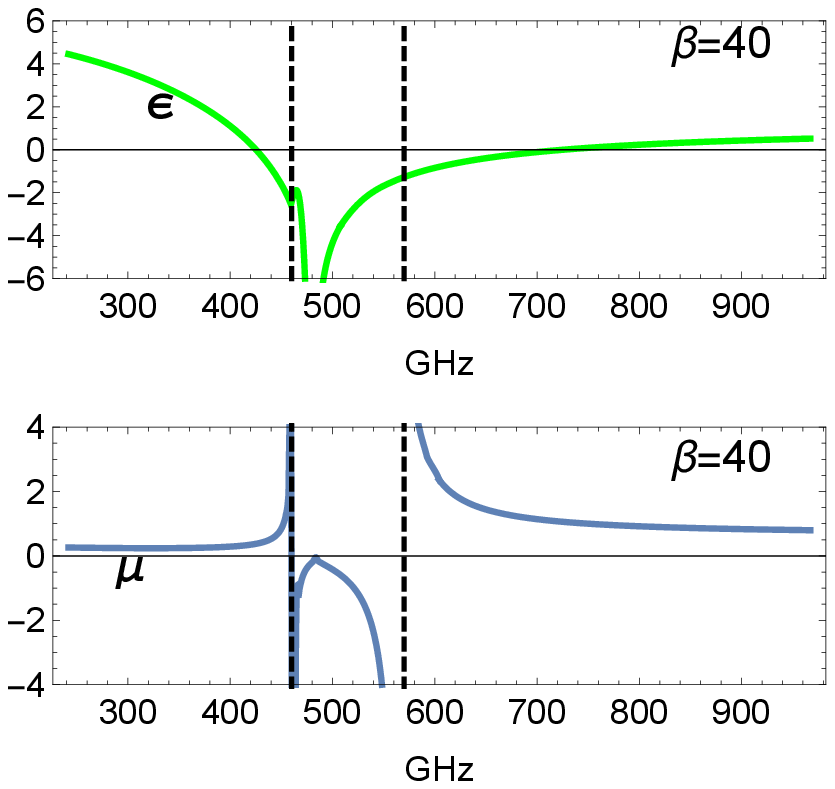}}
\hspace{0.1cm}
\subfigure[ ]{
\includegraphics[scale=0.48]{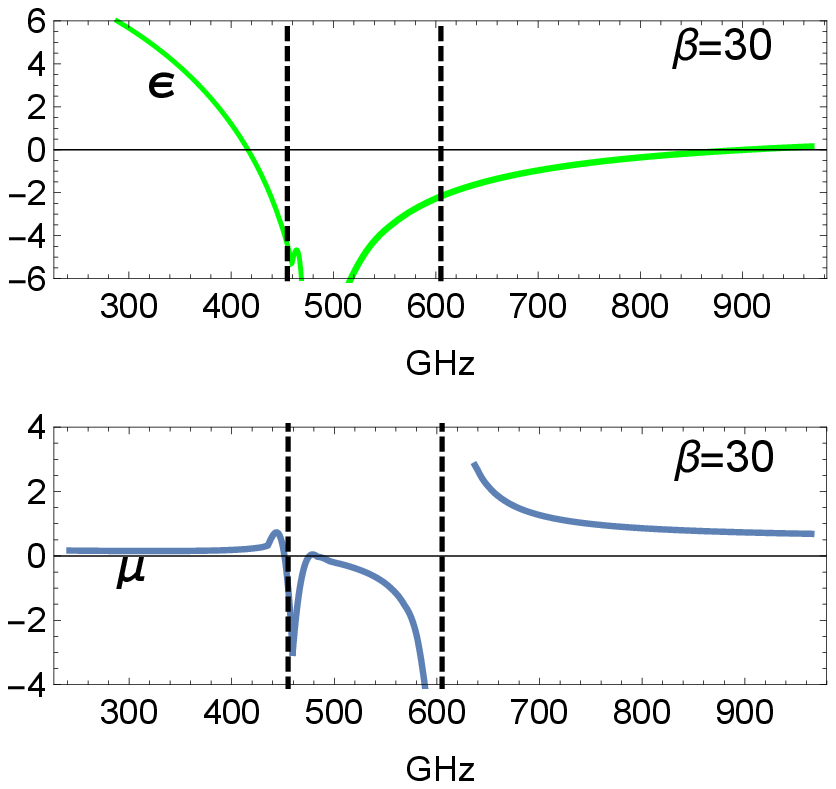}}
\caption{  We choose $k v_F=2.0 $ meV for the convenience of demonstration. (a) At $T=0$, the $\epsilon$ is positive in the frequency range of interest. The condition of NRI cannot be satisfied.  (b-d) are the frequency responses of $\epsilon$ and $\mu$ at different temperatures. It clearly shows that the range of NRI expands when $T$ increases. }
\label{freqresponse}
\end{figure}

\begin{figure}[t]
\centering
\subfigure[ ]{
\includegraphics[scale=0.41]{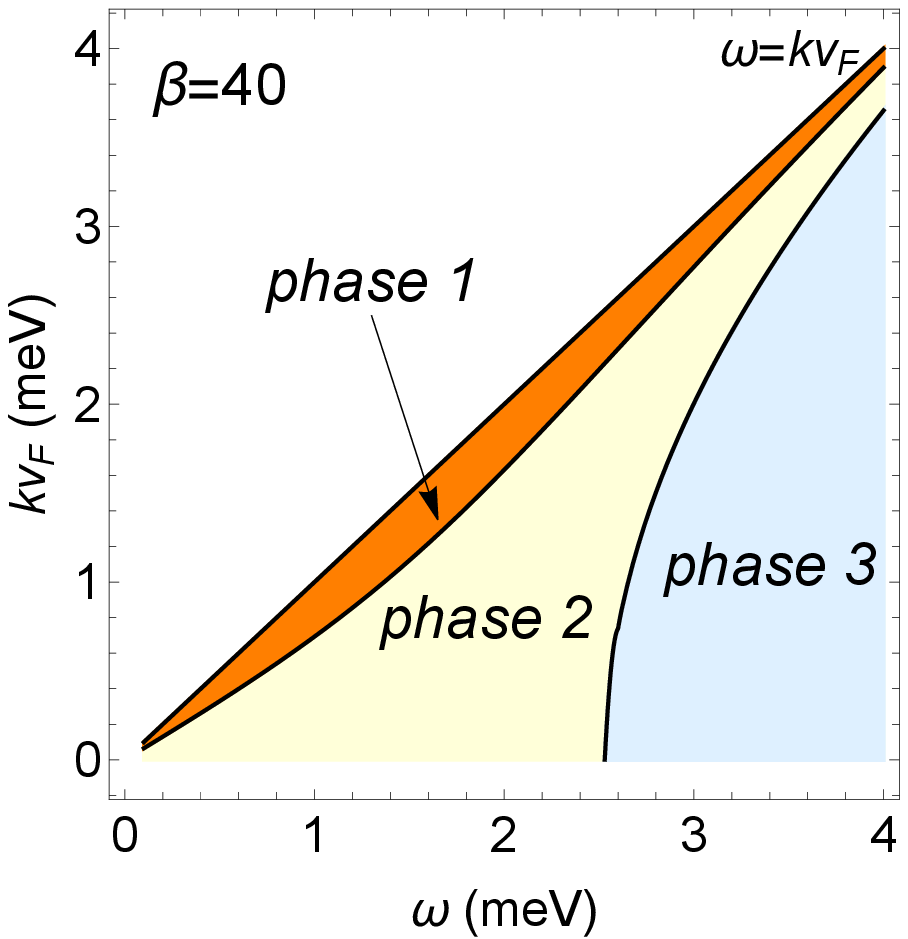}}
\hspace{0.1cm}
\subfigure[ ]{
\includegraphics[scale=0.47]{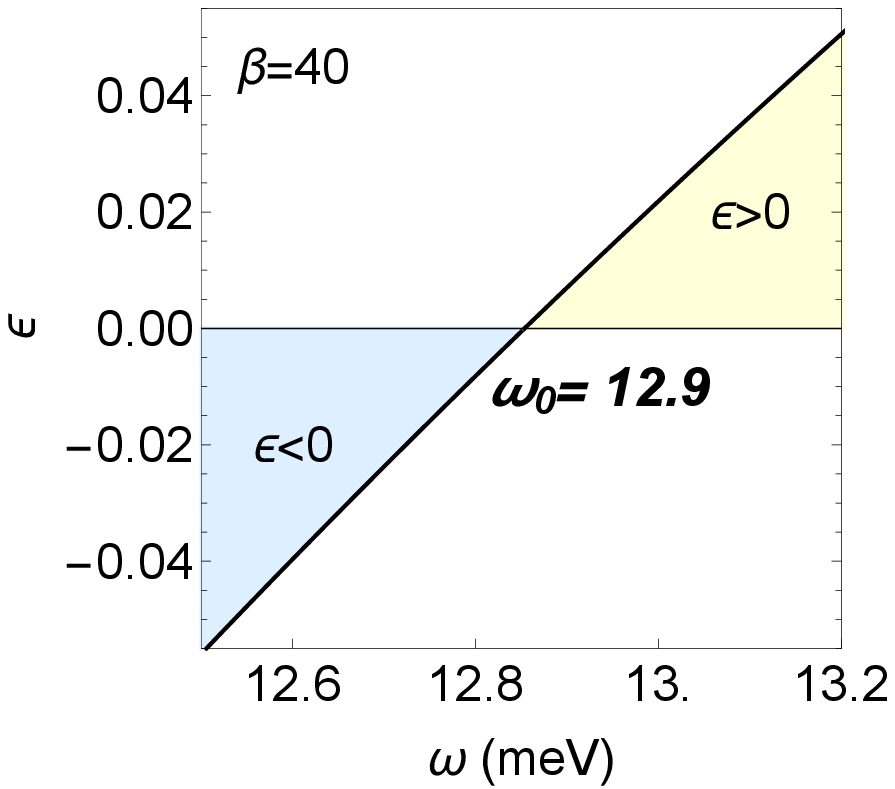}}
\caption{(a) Near room temperature, the phase diagram of phase 1(NRI), and the other two phases. These phases are defined by the signs of $\epsilon$ and $\mu$. (b) For $\omega>\omega_{0}$ ($\approx 12.9$ meV), the $\epsilon$ becomes positive. $\omega_0$ denotes the endpoint of phase 1. }
\label{phasediagram}
\end{figure}

 
%%%%%%%
\textit{Introduction}
%%%%%%%
The theoretical prediction of the possibility of negative index materials (NIM) was first proposed by Veselago in 1968 \cite{Veselago}. Veselago demonstrated that without violating the Maxwell's equations, there is no reason to rule out the existence of materials which have negative electric permittivity $\epsilon$ and magnetic permeability $\mu$. In 2000, Smith and coworkers displayed in their pioneering work that it was possible to fabricate artificial materials with negative refractive index (NRI) \cite{Smith}.  In 2003, two experiments were performed which confirmed the existence of NRI systems, in support of Veselago's hypothesis. Parazzoli \textit{et al.} constructed a wedge sample that was appropriate for free-space measurements \cite{Parazzoli}. Based on a ring and wire structure, the sample in Ref. \cite{Parazzoli} clearly illustrated NRI. Right after Parazzoli's work, Houck and colleagues successfully implemented NIM at certain frequency range by using a planar waveguide configuration to map the field pattern of microwaves that transmitted through wedge-shaped samples \cite{Houck}. More recent works have proposed that NIM may happen in chiral materials \cite{Zhang, Ukhtary, Hayata}. However, up to now there is no experimental report of NRI in natural materials \cite{Ramakrishna, Liu}. \\
\indent
In order to seek out NRI in natural materials, we investigate massless electron system which are rarely studied for physical system and may contain unexpected electromagnetic effects. It is untill the discovery of Dirac semimetals (DSM) that massless electrons find their positions in condensed matter systems. A physical system is typically surrounded in a finite temperature (FT) environment, therefore we study the finite temperature correction (FTC) of the massless electron system, following the methodology in the study of massive particle system \cite{Donoghue, Donoghue2}.
Zero effective mass theory at the Lagrangian level is scaleless by and in itself, and this leaves the temperature as the sole nominal energy scale. As there is no nominal scale incurred by mass, physical effects, no matter how small, are not necessarily overwhelmed by effects of which size is related to the mass if mass were existing. Therefore, it will be interesting to investigate the physical properties of massless electrons at FT. The DSM, in which electrons have zero effective mass and already exhibit intriguing optical properties such as inverse Farady effect \cite{IFE}, are prefect candidates, see \cite{DM} and the references therein. \\
\indent
In condensed matter systems, the electron behaviors are affected by numerous mechanisms such as the potential of lattice structures and electron-electron interactions. In DSM, all of the aforementioned effects result in zero band gap and linear dispersion near the Dirac cones \cite{DM}, and the material properties are consolidated in $v_F$, the Fermi velocity. The electromagnetic properties induced by electrons can best be studied by calculating vacuum polarization (VP), of which links with $\epsilon$ and $\mu$ were given by Weldon's article \cite{Weldon} and the successive work \cite{AM}. The VP is the first-order correction of the photon properties by photon-electron interaction. It describes a virtual process in which a photon produces an electron-hole pair and the pair recombines to give off a photon. The fact that the electrons and holes are massless has a profound effect on photon behaviors. We found that the VP contains a term with logarithmic divergence at $\omega\approx kv_F$. It comes from the resonance between the incident light and the electrons on Dirac cones. As the electrons are massless and have linear dispersion, the conditions of resonance (energy and momentum conservation) can be met by numerous number of states. These initial states can have energy and momentum $-a(\omega, \mathbf{k})$ where $0 < a < 1$ is an arbitrary constant. Those of the final state after absorbing a photon will be $(1-a)(\omega, \mathbf{k})$. Hence, the resonance becomes prominent. In contrast, for electrons with mass gaps or gapless electrons without linear dispersion (with energy form such as $p^2/2m$), only in specific states they can have resonance with an incoming photon. It is this divergence which makes the negative $\epsilon$ and $\mu$ possible. Thus the DSM that have Dirac cones can have NRI. In this article, we study the FTC to vacuum polarization of massless quantum electrodynamics (QED). As a result of the competitions among zero temperature (ZT) quantum loop effect and the FTC associated with it, the frequency and temperature ranges of NRI are identified, as will be shown below. \\
\indent
%%%%%%% section Vacuum polarization%%%%%%%%%%%%
\textit{Vacuum Polarization}
%%%%%%%%%%%%%%%%%%%%%%%%%%%%%%%%%
In DSM, the electrons are massless and have linear dispersion and constant velocity, denoted as $v_{F}$, the Fermi velocity. The Lagrangian density of electrons carrying charge $e$ reads 
\begin{equation}\label{lag}
\mathcal{L}= i \overline{\psi}  \left[  \gamma^{0}\partial_{t} + v_{F} \gamma^{j}\partial_{j} \right] \psi - e  \overline{\psi} \gamma^{\mu}A_{\mu} \psi ,
\end{equation}
where $\hbar$ is set to 1. The VP shall be calculated as it is closely related to electric permittivity $\epsilon$ and magnetic permeability $\mu$.
For a system in contact with a heat bath at rest, the real-time electron propagator is 
\begin{equation}\label{propagator}
S_{F}(p)=\frac{\slashed{p}+m}{p^2-m^2+i\eta}+2\pi i \frac{(\slashed{p}+m )\delta(p^2-m^2)}{e^{\beta|p\cdot u|}+1} , 
\end{equation}
where $u$ is the four-vector velocity of the system and $\slashed{p}$ denotes $\gamma^{\mu}p_{\mu}$ \cite{Fetter}. This is a general form of Lagrangian for arbitrary mass value. We will set $m=0$ for DSM when we start to evaluate VP. It is evident that $u=(1,0,0,0)$ in the rest frame of the heat bath. The appearance of $p$ implies that the poles of the propagator are at $p^{2}=m^2 v^{4}_{F} $ or $p^0=\pm v_{F} \sqrt{\mathbf{p}^2+m^2 v^2_{F}}$ in general. For massless electrons, we have $p^0=\pm |\mathbf p | v_F$.\\
\indent
We first consider the contribution of free electrons to the $\epsilon$ and $\mu$ by calculating VP tensor. The contribution arising from lattice shall be dealt with later. VP tensor is the integral of product of two massless electron propagators 
\begin{equation}\label{polarization}
\begin{aligned}
\pi_{\mu\nu}(k,& \omega )=i e^2\int \frac{d^4 p}{(2\pi)^4} \text{Tr}[\gamma_{\mu} (\slashed{p}+\slashed{K}) \gamma_{\nu}  (\slashed{p})]  \left[ \frac{1}{(p+K)^2} \right. \\
& \left.+ 2\pi i \frac{ \delta((p+K)^2) }{{e^{\beta|(p+K)\cdot u|}+1}} \right]  \left[ \frac{1}{p^2} + 2\pi i \frac{\delta(p^2)}{{e^{\beta |p\cdot u|}+1}} \right],
\end{aligned}
\end{equation}
where $K=(\omega, v_F \mathbf{k} )$ is the photon 4-momentum of the incident light and $K^2=\omega^2-v_F^2 k^2$.
Following Ref. \cite{Weldon}, Eq. (\ref{polarization}) can be decomposed into
\begin{equation}\label{eqpimunu}
\pi_{\mu\nu} (k,\omega)= \pi_{T}(k,\omega) P_{\mu\nu} +  \pi_{L}(k,\omega) Q_{\mu\nu},
\end{equation}
where $\pi_L$ and $\pi_T$ are respectively the longitudinal and transverse part of the polarization form factors.
The definitions of $P_{\mu \nu}$ and $Q_{\mu \nu}$ are given by
\begin{equation}
\begin{aligned}
P_{\mu \nu}& \equiv g_{\mu \nu}-u_{\mu} u_{\nu}+\frac{K_{\mu} K_{\nu}}{v_F^2 k^2}, \\
Q_{\mu \nu}& \equiv - \frac{1}{K^2 v_F^2 k^2}(v_F^2 k^2 u_{\mu} +\omega \tilde{K}_{\mu}) (v_F^2 k^2 u_{\nu} +\omega \tilde{K}_{\nu}), 
\end{aligned}
\end{equation}
where $\tilde{K}_{\mu} \equiv  K_{\mu} -\omega u_{\mu}$. $\pi_{\mu\nu} (k,\omega) $ is symmetric under the exchange of $\mu \leftrightarrow \nu$ and the VP should obey the gauge condition $ K^{\mu} \pi_{\mu \nu}=0$.
The relations between $\pi_{\mu\nu}$, $\pi_{L}$ and $\pi_{T}$ are
 \begin{equation}
 \pi_{L}(k,\omega) = \frac{-K^2 u^{\mu} u^{\nu} \pi_{\mu \nu} }{v_F^2 k^2},\quad \pi_{T}(k,\omega)= \frac{ -\pi_{L} + g^{\mu\nu}\pi_{\mu\nu} }{2}.
 \end{equation}
in which the FTC part of longitudinal polarization in massless electrons is 
\begin{equation}\label{polarizationTemp}
\begin{aligned}
&\pi_{L, \beta}(k,\omega)=\frac{-4e^2}{(2\pi)^3}\left( 1-\frac{\omega^2}{v^2_F k^2}\right) \int d^4 p \left[ \frac{\delta(p^2)}{(p+K)^2} f(p)\right. \\
&\left.+ 2\frac{\delta((p+K)^2)}{p^2} f ( p+K) \right] [2 (p^0+K^0)p^0 -p\cdot(p+K)] ,
\end{aligned}
\end{equation}
where $f(p)=1/(e^{\beta |p\cdot u|}+1)$ is Fermi-Dirac distribution function and the subscript $\beta$ denotes the FT part of the one-loop effect. 
The calculation can be advanced by expanding $f(p)$ in power series of $\exp(-\beta p^0)$. \\
%%%%%%%%%%%%%%%%%%%%%%%%%%%%%%%%%%
\indent
The power series of $\exp(-\beta p^0)$ converges fast and one can derive that the temperature dependence is proportional to $T^4$ when $T\rightarrow 0$. More specifically, we have
\begin{equation}
 \pi_{L, \beta}\sim\frac{-56\alpha \pi^3}{45K^2}T^4 , \quad
\pi_{T, \beta}\sim\frac{ -56\alpha \pi^3 }{45  K^2 } \left( 1+ \frac{\omega^2}{v^2_Fk^2} \right) T^4,
\end{equation}
where $\alpha$ is the fine structure constant in QED. In general, the FTC to loop amplitude will be in form of even power of $T$, as is evident from the Sommerfield expansion \cite{sommerfield}. Typically, the leading FTC of a loop amplitude will start from $T^2$ in a massive electron theory. In the current massless case, however, the leading FTC starts only from $T^4$. It is the result of cancellation of $T^2$ terms between the contributions of two electron propagators of like forms. \\
\indent
In principle, a $T^4$ FTC is negligibly small if the system exists certain energy scales such as electron mass. Typically the energy scale of a system is defined by the trace of energy-momentum tensor (EMT). In a massless theory, there is no intrinsic energy scale at the Lagrangian level as the trace of EMT of Lagrangian is set by the mass and $m=0$ for a massless theory. Therefore, it appears that the $T^{4}$ FTC has a chance to become manifest. But in a massless theory there will be an energy scale induced by the ZT VP \cite{Donoghue2}. The energy scale so induced is dubbed trace anomaly as massless theory originally has zero trace of EMT, and the VP tensor gives rise to non-zero trace; it defines the energy scale of the system including one-loop effects. Noted that at the same one-loop level electron self-energy diagram will induce a mass-correction term \cite{Weldon, Donoghue}. It seems that this effect may also exhibit a new energy scale. However, this effect does not contribute to the trace of EMT at one-loop level. Even if it induces an energy scale, the effect will emerge only at higher-loop calculation and hence can be safely neglected in the one-loop study.\\
\indent
Despite of the energy scale emerging from the ZT VP, fortunately, the leading term for this effect is a one-loop amplitude and this allows FTC, which is also a one-loop effect and its size could vary by tuning $\omega$ and $k$, to compete with the ZT VP effect. We conclude that the $T^4$ dependence of FTC will manifest only in a massless electron theory.\\
\indent
%%%%%%%%%%%%%%%%% Section NIM%%%%%%%%%%%%%%
\textit{Negative Refractive Index} 
%%%%%%%%%%%%%%%%%%%%%%%%%%%%%%%%%%%%%
We are going to show that there exists a fine-tuned region of frequency and wave number allowing the existence of negative $\epsilon$ and $\mu$ in the DSM at FT. We follow Eq. (2.26) in \cite{Weldon} which reads
\begin{equation}\label{epmu}
\begin{aligned}
 \epsilon(k,\omega) =   1- \frac{\pi_{L}}{K^2},  \quad
 \frac{1}{\mu(k,\omega)} = 1+ \frac{K^2 \pi_{T} - \omega^2 \pi_{L}}{v_F^2 k^2 K^2}.
 \end{aligned}
 \end{equation}
The full expression $\pi_{L(T)}$ is $\pi_{L(T)}=\pi_0 + \pi_{L(T), \beta}$, where $\pi_0$ is the ZT part of VP amplitude. 
The signs of $\epsilon$ and $\mu$, which give drastically different physical properties \cite{Liu}, depend on both $\omega$ and $k$.\\
\indent
The ZT part of the VP amplitude $\pi_0$ is given by
\begin{equation}\label{zeroPi}
\pi_0(k,\omega)=\frac{\alpha}{3 \pi}(\omega^2-k^2 v^{2}_{F})\ln\left( \frac{\omega^2-k^2 v^{2}_{F} }{\mu_{m}^2} \right),
\end{equation}
where $\mu_{m}$ is the infrared cut-off energy \cite{Donoghue2}. We choose the cut-off energy $\mu_{m}$ being $\text{cm}^{-1}$ ( $\approx 0.1$ meV), depending on the size of samples. The physical reason is that the infrared cut-off is to cut off soft photons, which means the low energy ones. The above ZT properties can be found in the ZT part of $\epsilon$, shown in FIG. \ref{freqresponse} (a). If only the ZT part of the VP is considered, both $\epsilon$ and $\mu$ cannot have negative value. This can be seen easily by substituting Eq. (\ref{zeroPi}) into Eq. (\ref{epmu}) with $K^2$ being greater than the infrared cut-off. It is by definition true because the infrared cutoff is the resolution limit of measurements. Furthermore, the numerical result is not sensitive to the choice value of $\mu_m$. \\
\indent
The equations in Eq. (\ref{epmu}) were calculated numerically with the energy range within a few meV apart from $k=0$ since DSM show linear dispersion relation only near the Dirac point. The most interesting part is the existence of NIM, for which the $\epsilon$ and $\mu$ are both negative. The plots of $\epsilon$ and $\mu$ with respect to incident light frequency $\omega$ under different temperatures are shown in FIG. \ref{freqresponse} (b-d). The region marked between two dashed lines is where NRI takes place, which becomes larger when temperature raises. The fact shows that temperature plays a crucial role in the occurrence of NRI. One can also see that the product $\epsilon \mu \geq 1$, however, the refractive index $n=\pm\sqrt{\epsilon \mu}$ should take a negative sign to preserve causality \cite{Veselago}.  \\
\indent
FIG. \ref{phasediagram} (a) gives the phase diagram of DSM where phase 1 ($\epsilon<0$, $\mu<0$), phase 2 ($\epsilon<0$, $\mu>0$) and phase 3 ($\epsilon>0$, $\mu>0$) near room temperature are shown. Recall that at ZT the system exhibits no NRI. We have calculated the range where negative $\epsilon$ is allowed numerically. The accuracy of $\delta=\omega-k v_{F}$ is $10^{-7}$ meV. We found that for $\omega>\omega_{0}\approx12.9$ meV, $\epsilon$ has no chance to be negative, see FIG. \ref{phasediagram} (b).\\
%%%%%%%%%%%%%%%%%% FIG contour plots %%%%%%%%%%%%%%
%%%%%%%%%%%%%%%%%%%%%%%%%%%%%%%%%%%%%%%%%
%%%%%%%%%%%%%%%%%%FIG phase diagram%%%%%%%%%%%%%%%
%%%%%% %%%%%%%%%%%%%%%%%%%%%%%%%%%%%%%%%
\indent
We now express analytically the NIM property of massless electrons. When $k v_F$ is close to $\omega$, we can derive the asymptotic behavior of permittivity which reads
\begin{equation}\label{epsilonApprox}
\epsilon \approx 1 + \mathcal{C}_{\omega, k}  + \frac{2\pi^3 \alpha}{3 k^2  v_F^2 \beta^2}  \ln\frac{\beta(\omega-k v_F)}{2} \qquad (k v_F\rightarrow\omega),
\end{equation}
where the symbol $\mathcal{C}_{\omega, k} $ is a  $\omega$ and $k$ dependent summation but converges to a finite value. Detailed derivation of Eq. (\ref{epsilonApprox}) can be found in Supplemental Material. Due to the logarithmic term $\epsilon$ can be negative and large in magnitude when $k v_F$ is very close to $\omega$. We can also derive that $\pi_{T, \beta} \approx 2\alpha \pi/3\beta^2$ when $ k v_F\approx \omega$. By Eq.  (\ref{epmu}) and (\ref{epsilonApprox}), 
\begin{equation}\label{muApprox}
\frac{1}{\mu}
\approx 1+\mathcal{C}_{\omega, k}  + \frac{2\pi\alpha}{3k^2 v_F^2\beta^2}+ \frac{2\pi^3 \alpha}{3 k^2 v_F^2 \beta^2} \ln\frac{\beta( \omega-k v_F)}{2}.
\end{equation}
From Eq. (\ref{epsilonApprox}) and (\ref{muApprox}), we conclude that both $\epsilon$ and $\mu$ can be negative when $k v_F\approx \omega$.\\
\indent
Above results come from the VP of free electrons. We now apply our calculation to realistic systems. Taking $\text{Cd}_3 \text{As}_2 $ as an example, Eq. (\ref{epsilonApprox}) has to be modified to account for the lattice contribution to the permittivity:
%%%%%%%%%%%%%%%%%%%%%FIG with lattice effect%%%%%%%%%%%%%%%%%%%%
\begin{figure}[t]
\centering
\subfigure[ ]{
\includegraphics[scale=0.51]{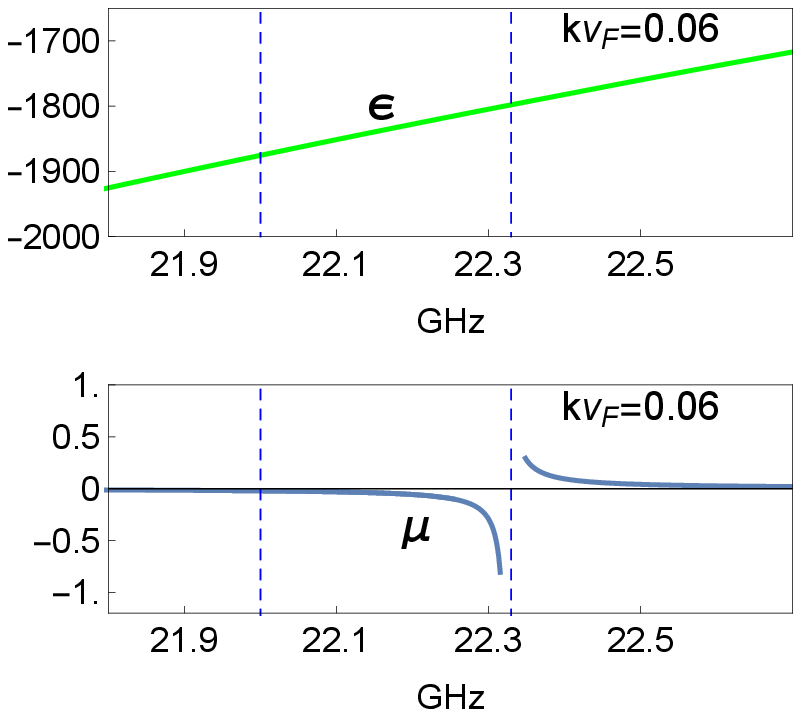}}
\hspace{0.1cm}
\subfigure[ ]{
\includegraphics[scale=0.49]{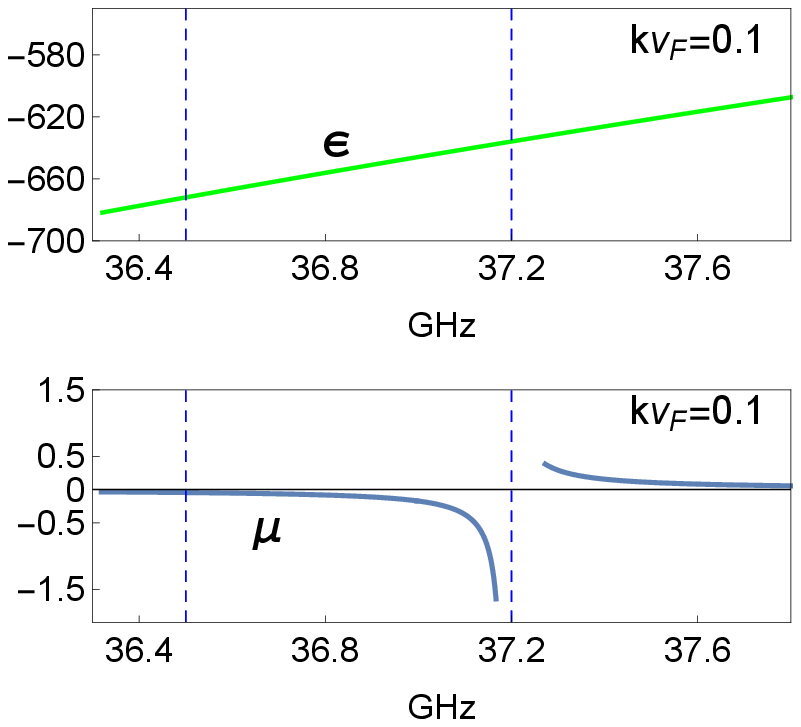}}
\hspace{0.1cm}
\subfigure[ ]{
\includegraphics[scale=0.44]{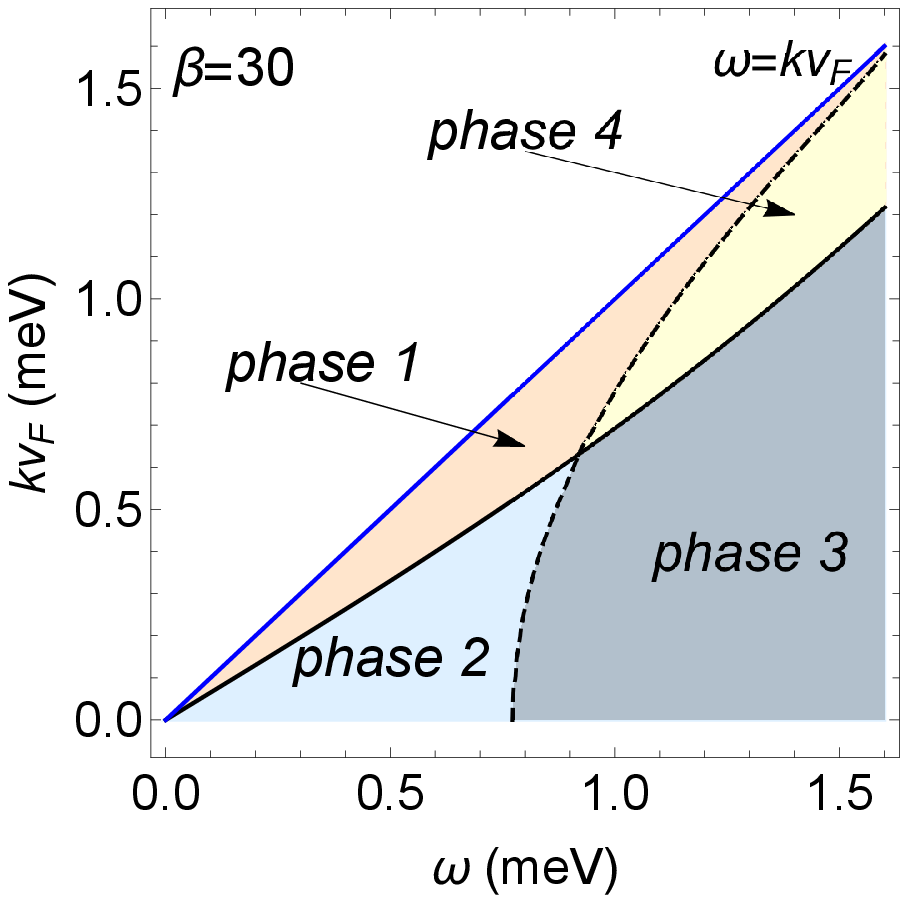}}
\hspace{0.1cm}
\subfigure[ ]{
\includegraphics[scale=0.45]{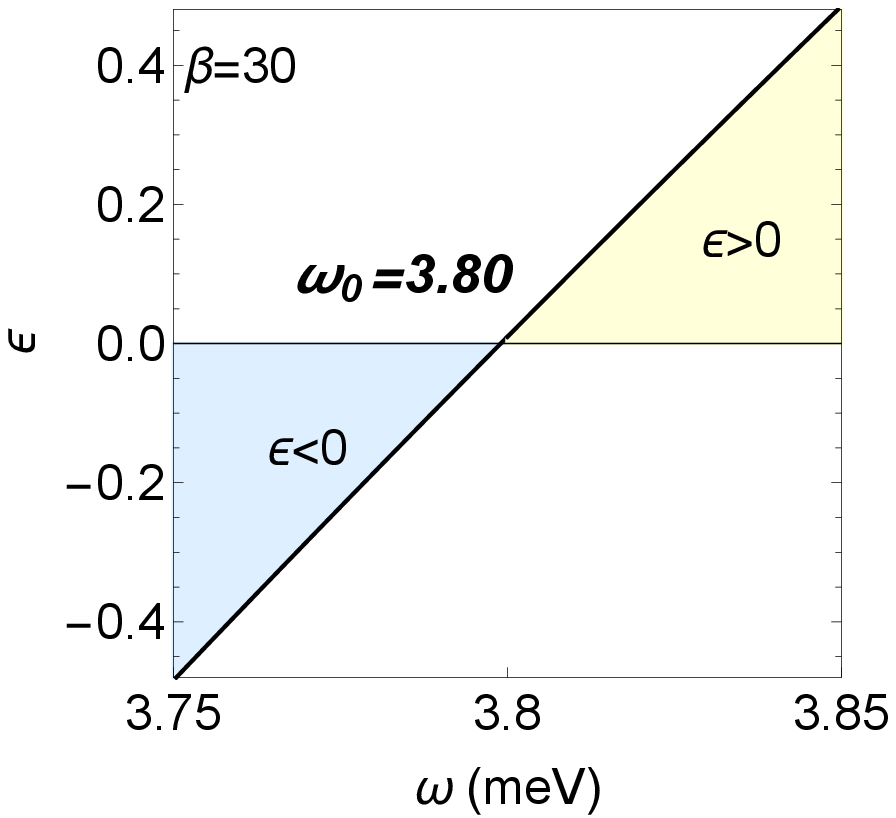}}
\caption{ With lattice contribution $\epsilon_{L}=17$ \cite{Cd3As2} and temperature $\beta=30$ $\text{eV}^{-1}$: (a) and (b) are the plots of frequency responses of $\epsilon$ and $\mu$ when $kv_F=0.06$ and $0.1$ meV, respectively. (c) The phase diagram of phase 1, phase 2, phase 3 and phase 4 ($\epsilon>0$, $\mu<0$). Left (right) of the dashed line shows the sign boundary of $\epsilon$ which is smaller (greater) than zero. (d) For $\omega>\omega_{0}$ ($\approx 3.80$ meV), the $\epsilon$ becomes positive.}
\label{freqresponsetLattice}
\end{figure}
%%%%%%%%%%%%%%%%%%%%%%%%%%%%%%%%%%%%%%%%%%
\begin{equation}\label{epsilonLattice}
\epsilon \approx \epsilon_{L} + \mathcal{C}_{\omega, k} + \frac{2\pi^3 \alpha}{3 k^2  v_F^2 \beta^2}  \ln\frac{\beta(\omega-k v_F)}{2} \qquad (k v_F\rightarrow\omega),
\end{equation}
where $\epsilon_{L}$ contains the lattice effects. Experiments showed that  $\epsilon_{L} \approx 17$ for $\text{Cd}_3 \text{As}_2 $ \cite{Cd3As2}.
On the other hand, the permeability is modified very little by the lattice effects such as phonons and high frequency polarizations and hence needs not to be revised. The resulting NRI regions become smaller than that of free electrons. They are plotted in Fig. \ref{freqresponsetLattice} at $\beta=30$ $\text{eV}^{-1}$ for $\epsilon$, $\mu$, phase diagram and the critical region. One novelty is now a new phase appears (phase 4) due to the shift of the $\epsilon$ boundary.\\
\indent   
Clearly the negativity of $\epsilon$ and $\mu$ and hence the NRI originate from the logarithmic divergence at $\omega\approx kv_F$. A massless electron with linear dispersion occupying a state in a Dirac cone can absorb a photon and make transition to a higher energy state in the Dirac cones, and hence, induce the resonance. Though the resonance can occur only if the $v_F$ is close to the light speed in the matter, once it occurs, there will be numerous states that can contribute. Therefore, the resonance is prominent. We thus conclude that the negative $\epsilon$ and $\mu$ is due to the very existence of Dirac cones, the unique band structure of DSM. 
The broadening phenomenon in FIG. \ref{freqresponse} is due to the spillover of Fermi-Dirac distribution function when temperature increases; the higher the temperature is, the more states become available for such a resonance. \\
\indent
As shown above, both the $\epsilon$ and $\mu$ have $\omega$ and $k$ dependencies. The latter dependency is often been referred to as the spatial dispersion or nonlocal effect. Since the relation between $\omega$ and $k$ of EM waves in media is determined by the $\epsilon$ and $\mu$. They have both dependencies naturally. However, we note that this is different from the nonlocal effect in the metamaterials \cite{Elser}. Due to the size of constitutive components and inhomogeneity of metamaterials, the $\epsilon$ and $\mu$ cannot be simply taken as averaged quantities. Instead, they have to be calculated to account for the nonlocal effect. This additional effort is not required in natural materials. Therefore, DSM can provide a straightforward experimental demonstration.\\
\indent
In order to realize NRI in real systems, one more requirement has to be satisfied. The light speed in the material is $\omega/k$. Both $\epsilon$ and $\mu$ are negative when $kv_F \approx  \omega$ and hence $c/\sqrt{| \epsilon \mu}|$ has to be close to $v_F$. Typically $v_F \approx  c/100$ in materials. Hence, $\sqrt{\epsilon\mu}$ has to be around 100, very large in magnitude. The requirement can always be fulfilled. In view our Eqs. (12) and (13), both $\epsilon$ and $\mu$ become negative and large only when $\omega \approx kv_F$. This is when the logarithmic function becomes negative and cancels the large and positive term $\mathcal{C}_{\omega,k}$. In principle the value of $\epsilon$ in Eq. (12) can be arbitrarily large. So does the value of $\mu$ if $1/\mu$ approaches zero. Hence, the requirement mentioned before can be satisfied as long as $\omega \approx kv_F$.\\
\indent
However, any system inevitably has dissipation. Its effect can be approximated by replacing $\omega$ with $\omega+i/\tau$ where $\tau$ is the relaxation time \cite{IFE}. This results in the factor $\ln|\beta (\omega - k v_F)/2|$ in our VP to be substituted by $\ln|\beta \sqrt{(\omega - k v_F )^2+1/\tau^2}/2|$. A recent experiment \cite{Qfactor} showed that the quality factor ($Q\approx 1/\omega\tau$) can be as high as $3.7$. As a result the value of the logarithmic function can still be negative and large in magnitude as long as the temperature is high enough. In view of Eq. (\ref{epsilonLattice}), we found that the $\epsilon$ (and also $\mu$) sat comfortably in the negative region if $\omega/k_BT \lessapprox 1/40$.\\
%%%%%%%%%%%%%%%%%%%%%%%%%%%%%%%%%%%%%%%
\indent
\textit{Conclusion}
We have calculated the VP of massless electrons at FT with the real-time propagators formalism. The analytic expression was obtained. Its asymptotic behavior at low temperature is proportional to $T^4$. Our calculation has been applied to $\text{Cd}_3 \text{As}_2$ to show the frequency region of NRI at FT. Both negative $\epsilon$ and $\mu$ arise from the logarithmic divergence at $kv_F\approx \omega $. Incident photons with such a dispersion relation facilitate the resonant excitation of electrons in the Dirac cones. This is similar to the situation in metamaterials where NRI is realized by approaching a certain resonance frequency from below. It is the very nature of electron being massless and with linear dispersion relation which makes DSM to be a natural material candidate exhibiting NRI. Lastly, we found that higher temperature favors the occurrence of NRI. This work is supported by Chung Yuan Christian University and Department of Physics, National Taiwan University.
%%%%%%%%%%%%%%%%%%%%%%%%%%%%%%%%%%%%%%%%

%%%%%%%%%%%%%%%%%%%%%%%%%%%%%%%%%%%%%%%%%%%

\end{document}